\documentclass[vecphys]{svmult}
\usepackage{makeidx}         
\usepackage{graphicx}        
\usepackage{multicol}        
\usepackage[bottom]{footmisc}
\makeindex             
\newcommand{\etal}{et\,al.\ }
\newcommand{\logg}{\mbox{$\log g$}}
\newcommand{\Teff}{\mbox{$T_\mathrm{eff}$}}
\newcommand{\gppr}{\stackrel{>}{\scriptstyle \sim}}
\newcommand{\gappr}{\raisebox{-0.4ex}{$\gppr $}}
\newcommand{\lppr}{\stackrel{<}{\scriptstyle \sim}}
\newcommand{\lappr}{\raisebox{-0.4ex}{$\lppr $}}

\newcommand{\msun}{\ensuremath{\, {\rm M}_\odot}}
\newcommand{\lsun}{\ensuremath{\,{\rm L}_\odot}} 

\newcommand{\ion}[2]{\mbox{#1\,{\sc #2}}}

\begin{document}

\title*{Element Abundance Determination in Hot Evolved Stars}
\author{Klaus Werner}
\institute{Institute for Astronomy and Astrophysics, Kepler Center for Astro
  and Particle Physics, University of T\"ubingen,
  Sand~1, 72076~T\"ubingen, Germany, 
\texttt{werner@astro.uni-tuebingen.de}}
\maketitle

\begin{abstract}
The hydrogen-deficiency in extremely hot post-AGB stars of spectral class PG1159
is probably caused by a (very) late helium-shell flash or a AGB final thermal
pulse that consumes the hydrogen envelope, exposing the usually-hidden
intershell region. Thus, the photospheric element abundances of these stars
allow us to draw conclusions about details of nuclear burning and mixing
processes in the precursor AGB stars. We compare predicted element abundances to
those determined by quantitative spectral analyses performed with advanced
non-LTE model atmospheres. A good qualitative and quantitative agreement is
found for many species (He, C, N, O, Ne, F, Si, Ar) but discrepancies for others
(P, S, Fe) point at shortcomings in stellar evolution models for AGB stars.
Almost all of the chemical trace elements in these hot stars can only be
identified in the UV spectral range. The \emph{Far Ultraviolet Spectroscopic
Explorer} and the \emph{Hubble Space Telescope} played a crucial role for this
research.
\end{abstract}

\section{Introduction}

The chemical evolution of the Universe is driven by the nucleosynthesis of elements
in stars. Evolved stars return a significant fraction of their mass 
to the interstellar medium. This matter is enriched with heavy elements which
were produced in the stellar interior and dredged-up to the surface by
convective motions. For quantitative modeling of galactic chemical evolution
it is crucial to know the stellar yields of chemical elements, i.e., how much
metals are produced by which stars. Yields are computed from stellar evolution
models, however, several uncertainties in modeling can strongly affect the
yields. Among the most serious problems are mixing processes (convection) and
some particular nuclear reaction rates. One solution to these problems is a
comparison of predicted surface abundances with observations. Quantitative
spectroscopy is therefore a powerful tool to calibrate free modeling
parameters, e.g., those associated with convective overshoot.

This paper is restricted to evolved low-mass stars, or to be more
precise, to hot post-AGB stars. About 95\% of all stars in our Galaxy end their
life as a white dwarf. These low- and intermediate-mass stars
($\approx$1--8\msun) roughly produce 50\% of the metal yields,
mainly during the phase of AGB evolution in combination with strong
radiatively-driven mass loss. We demonstrate here that quantitative
abundance analyses of particular elements in post-AGB stars provide valuable
insight into AGB-star nucleosynthesis processes.

We further confine our paper to very hot \emph{hydrogen-deficient}
post-AGB stars. The reason is that these particular objects offer the unique
possibility to directly access the nucleosynthesis products. 
Why do we concentrate on hot stars, i.e., those having
\Teff$\approx$100\,000\,K? This is because the spectra of cooler objects are
wind-contaminated. These are the Wolf-Rayet type central stars of planetary
nebulae and their atmospheres are much more difficult to model. In addition,
very weak spectral features from rare elements can be smeared out by atmospheric
motion and
become undetectable. We also exclude from our studies hot white dwarfs and subdwarf
O/B stars, because their nucleosynthesis history was wiped out by diffusion
effects in their atmospheres.

\section{Quantitative spectral analysis of PG1159 stars}

\begin{figure*}[bth]
\includegraphics[width=\textwidth]{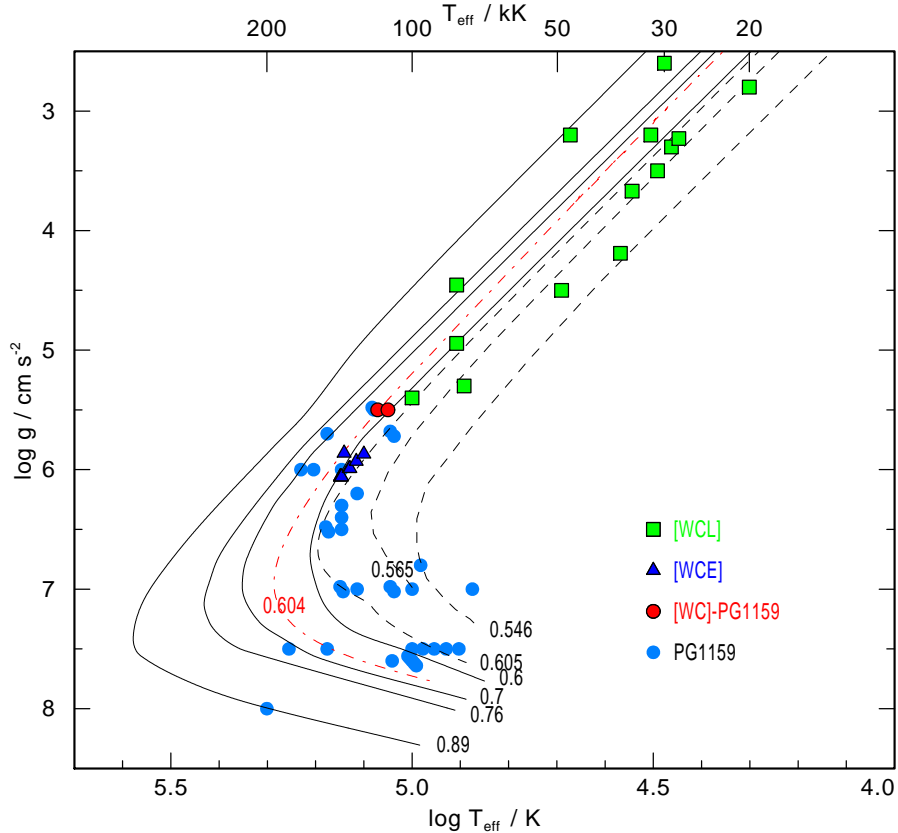}
\caption{Hot hydrogen-deficient post-AGB stars in the
$g$--\Teff--plane. We identify Wolf-Rayet central stars of early and
late type ([WCE], [WCL], from \cite{hamann:97}), PG1159 stars
(from \cite{werner:06}) as well as two [WC]--PG1159 transition objects
(Abell~30 and 78). Evolutionary tracks are from \cite{schoenberner:83}
and \cite{bloecker:95b} (dashed lines), \cite{wood:86} and
\cite{2003IAUS..209..111H} (dot-dashed line) (labels: mass in
M$_\odot$).  The latter $0.604\msun$ track is the final CSPN track
following a VLTP evolution and therefore has a H-deficient
composition.}\label{fighrd}
\end{figure*}

We report on our work on PG1159 stars, a group of 40 extremely hot hydrogen-deficient
post-AGB stars \cite{werner:06}.  Their effective temperatures (\Teff) range
between 75\,000--200\,000~K. Many of them are still heating up along
the constant-luminosity part of their post-AGB evolutionary path in the
HRD ($L \approx 10^4$\lsun) but most of them are already fading along
the hot end of the white dwarf cooling sequence (with $L$ $\gappr$
10\,\lsun). Luminosities and masses are inferred from spectroscopically
determined \Teff\ and surface gravity (\logg) by comparison with
theoretical evolutionary tracks. The position of analysed PG1159 stars
in the ``observational HR diagram'', i.e. the \Teff--\logg\ diagram, are
displayed in Fig.\,\ref{fighrd}. The high-luminosity stars have low
\logg\ ($\approx$\,5.5) while the low-luminosity stars have a high
surface gravity ($\approx$\,7.5) that is typical for white dwarf (WD)
stars. The derived masses of PG1159 stars have a mean of 0.62\,\msun, a
value that is practically identical to the mean mass of WDs. The PG1159
stars co-exist with hot central stars of planetary nebulae and the
hottest hydrogen-rich (DA) white dwarfs in the same region of the
HR diagram. About every other PG1159 star is surrounded by an old, extended
planetary nebula.

What is the characteristic feature that discerns PG1159 stars from
``usual'' hot central stars and hot WDs? Spectroscopically, it is the
lack of hydrogen Balmer lines, pointing at a H-deficient surface
chemistry. The proof of H-deficiency, however, is not easy: The stars
are very hot, H is strongly ionized and the lack of Balmer lines could
simply be an ionisation effect. In addition, every Balmer line is
blended by a Pickering line of ionized helium. Hence, only detailed
modeling of the spectra can give reliable results on the photospheric
composition. The high effective temperatures require non-LTE modeling of
the atmospheres. Such models for H-deficient compositions have only
become available in the early 1990s after new numerical techniques have
been developed and computers became capable enough.

\begin{figure*}[bth]
\includegraphics[width=\textwidth]{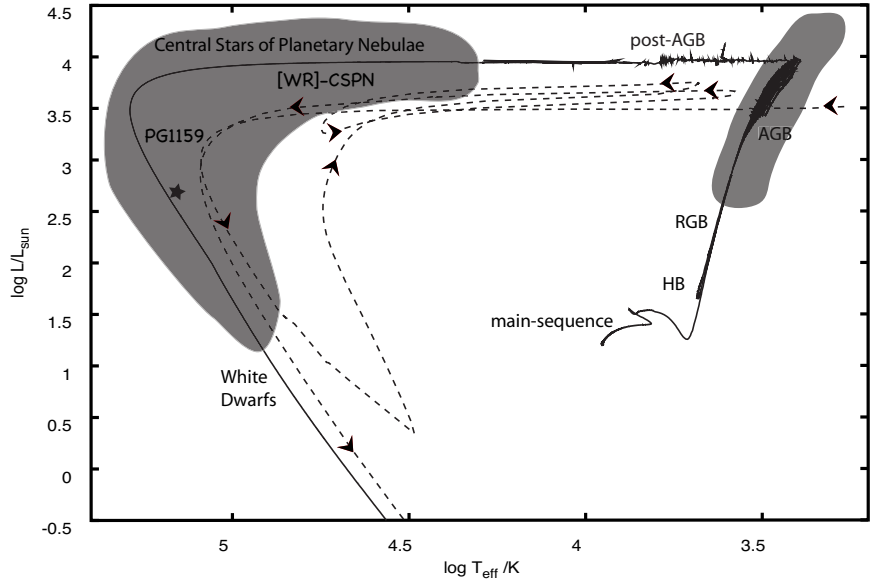}
\caption{\label{fig:hrd} Complete stellar evolution track with an initial mass
of $2\msun$ from the main sequence through the RGB phase, the HB to the
AGB phase, and finally through the post-AGB phase that includes the
central stars of planetary nebulae to the final WD stage. The solid line
represents the evolution of a H-normal post-AGB star. The dashed line
shows a born-again evolution of the same mass, triggered by a very late
thermal pulse, however, shifted by approximately $\Delta \log \Teff = -
0.2$ and $\Delta \log L/\lsun = - 0.5$ for clarity. The ``$\star$'' shows the
position of PG1159-035 \cite{werner:06}.
}
\end{figure*}                                                                                        

The first quantitative spectral analyses of optical spectra from PG1159
stars indeed confirmed their H-deficient nature \cite{werner:91}. It
could be shown that the main atmospheric constituents are C, He, and
O. The typical abundance pattern is C=0.50, He=0.35, O=0.15 (mass
fractions). It was speculated that these stars exhibit intershell matter
on their surface, however, the C and O abundances were much higher than
predicted from stellar evolution models. It was further speculated that
the H-deficiency is caused by a late He-shell flash, suffered by the
star during post-AGB evolution, laying bare the intershell layers. The
re-ignition of He-shell burning brings the star back onto the AGB,
giving rise to the designation ``born-again'' AGB star
\cite{iben:83a}. If this scenario is true, then the intershell
abundances in the models have to be brought into agreement with
observations. By introducing a more effective overshoot prescription for
the He-shell flash convection during thermal pulses on the AGB,
dredge-up of carbon and oxygen into the intershell can achieve this
agreement \cite{herwig:99c}. Another strong support for the born-again
scenario was the detection of neon lines in optical spectra of some
PG1159 stars \cite{werner:94}. The abundance analysis revealed Ne=0.02,
which is in good agreement with the Ne intershell abundance in the
improved stellar models.

If we accept the hypothesis that PG1159 stars display former
intershell matter on their surface, then we can in turn use these stars
as a tool to investigate intershell abundances of other
elements. Therefore, these stars offer the unique possibility to directly
see the outcome of nuclear reactions and mixing processes in the
intershell of AGB stars. Usually the intershell is kept hidden below a
thick H-rich stellar mantle and the only chance to obtain information
about intershell processes is the occurrence of the 3rd
dredge-up. This indirect view of intershell abundances makes the
interpretation of the nuclear and mixing processes difficult,
because the abundances of the dredged-up elements may have been changed
by additional burning and mixing processes in the H-envelope (e.g.,
hot-bottom burning). In addition, stars with an initial mass below
1.5~\msun\ do not experience a 3rd dredge-up at all.

We note that the central stars of planetary nebulae of
spectral type [WC] are believed to be immediate progenitors of PG1159
stars, representing the evolutionary phase between the early post-AGB
and PG1159 stages. This is based on spectral analyses of [WC] stars
which yield very similar abundance results \cite{hamann:97}. We do not
discuss the [WC] stars here because the analyses of trace elements are
much more difficult or even impossible due to strong line broadening in
their rapidly expanding atmospheres.

\section{Three different late He-shell flash scenarios}

The course of events after the final He-shell flash is qualitatively
different depending on the moment when the flash starts. We speak about
a very late thermal pulse (VLTP) when it occurs in a WD, i.e. the star
has turned around the ``knee'' in the HR diagram and H-shell burning has
already stopped (Fig.\,\ref{fig:hrd}). The star expands and develops a
H-envelope convection zone that eventually reaches deep enough that
H-burning sets in (a so-called hydrogen-ingestion flash). Hence H is
destroyed and whatever H abundance remains, it will probably be shed
from the star during the ``born-again'' AGB phase. A late thermal pulse
(LTP) denotes the occurrence of the final flash in a post-AGB star that
is still burning hydrogen, i.e., it is on the horizontal part of the
post-AGB track, before the ``knee''. In contrast to the VLTP case, the
bottom of the developing H-envelope convection zone does not reach deep
enough layers to burn H. The H-envelope (having a mass of about
$10^{-4}$~\msun) is mixed with a few times $10^{-3}$~\msun\ intershell
material, leading to a dilution of H down to about H=0.02, which is
below the spectroscopic detection limit. If the final flash occurs
immediately before the star departs from the AGB, then we talk about an
AFTP (AGB final thermal pulse). In contrast to an ordinary AGB thermal
pulse the H-envelope mass is particularly small. Like in the LTP case, H
is just diluted with intershell material and not burned. The remaining H
abundance is relatively high, well above the detection limit (H $\gappr$
0.1).

There are three objects, from which we believe to have witnessed a
(very) late thermal pulse during the last $\approx$\,100 years. FG~Sge
suffered a late flash in 1894 \cite{gonzalez:98}. The star became rich
in C and rare earth elements. It most probably was hit by an LTP, not a
VLTP, because it turned H-deficient only recently (if at all, this is
still under debate). As of today, FG~Sge is located on or close to the
AGB.

V605~Aql experienced a VLTP in 1917 \cite{clayton:97}. Since then,
it has quickly evolved back towards the AGB, began to reheat and is now
in its second post-AGB phase. It has now an effective temperature of the
order 100\,000~K and is H-deficient.

Sakurai's object (V4334~Sgr) also experienced a VLTP, starting
around 1993 \cite{duerbeck:96}. It quickly evolved back to the AGB and
became H-deficient. Recent observations indicate that the reheating of
the star already began, i.e., its second departure from the AGB might
just have begun.

The spectroscopic study of FG~Sge and Sakurai's object is particularly
interesting, because we can observe how the surface abundances change
with time. The stars are still cool, so that isotopic ratios can be
studied from molecule lines \cite{pavlenko:2004} and abundances of many
metals can be determined. The situation is less favorable with the hot
PG1159 stars: All elements are highly ionised and for many of them no
atomic data are available for quantitative analyses. On the other hand,
in the cool born-again stars the He-intershell material is once again
partially concealed.

\section{Comparison of observed and predicted abundances}

Abundance analyses of PG1159 stars are performed by detailed fits to
spectral line profiles. Because of the high \Teff\ all species are
highly ionized and, hence, most metals are only accessible by UV
spectroscopy. Optical spectra always exhibit lines from \ion{He}{ii} and
\ion{C}{iv}. Only the hottest PG1159 stars display additional lines of N,
O, and Ne (\ion{N}{v}, \ion{O}{vi}, \ion{Ne}{vii}). For all other species
we have utilized high-resolution UV spectra that were taken with the
\emph{Hubble Space Telescope} (HST) and the \emph{Far Ultraviolet Spectroscopic
Explorer} (FUSE). FUSE allowed observations in the Lyman-UV range
($\approx$\,900--1200~\AA) that is not accessible with HST, and this
turned out to be essential for most results reported here.

\begin{figure*}[bth]
\includegraphics[width=1.0\textwidth]{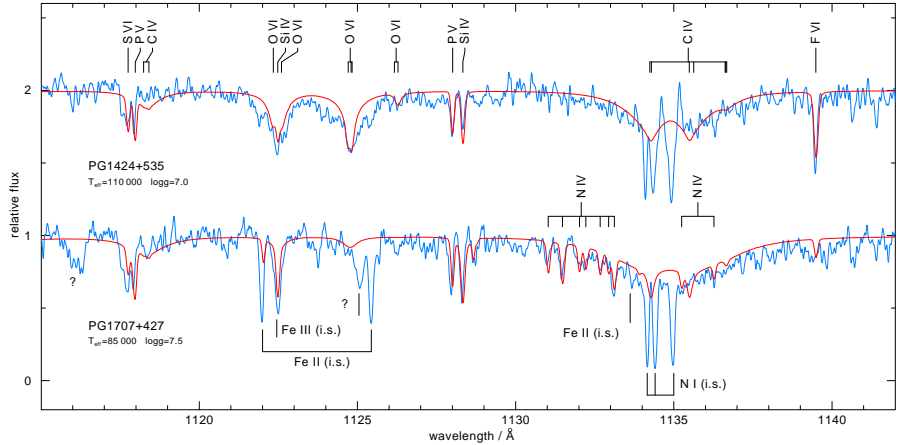}
\caption{\label{fig:psifn} Detail from FUSE spectra of two relatively
  cool PG1159 stars. Note the
  following features. The
  \ion{F}{vi}~1139.5~\AA\ line which is the first detection of F at all in
  a hot post-AGB star; the \ion{P}{v} resonance doublet at 1118.0 and
  1128.0~\AA, the first discovery of P in PG1159 stars; the
  \ion{N}{iv} multiplet at 1132~\AA. Also detected are lines from
  \ion{Si}{iv} and \ion{S}{vi}. The broader features stem from \ion{C}{iv}
  and \ion{O}{vi} \cite{reiff:07}.}
\end{figure*}

\begin{figure}[t]
\includegraphics[width=0.5\textwidth]{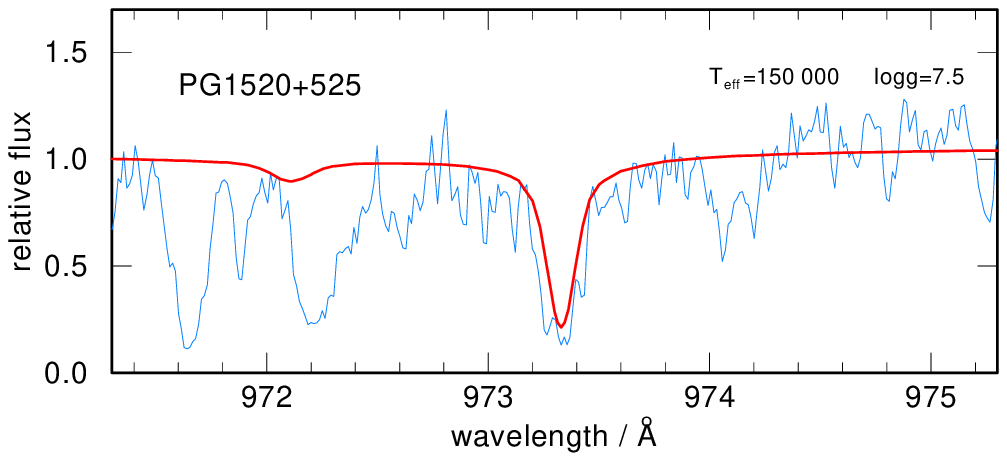}
\includegraphics[width=0.5\textwidth]{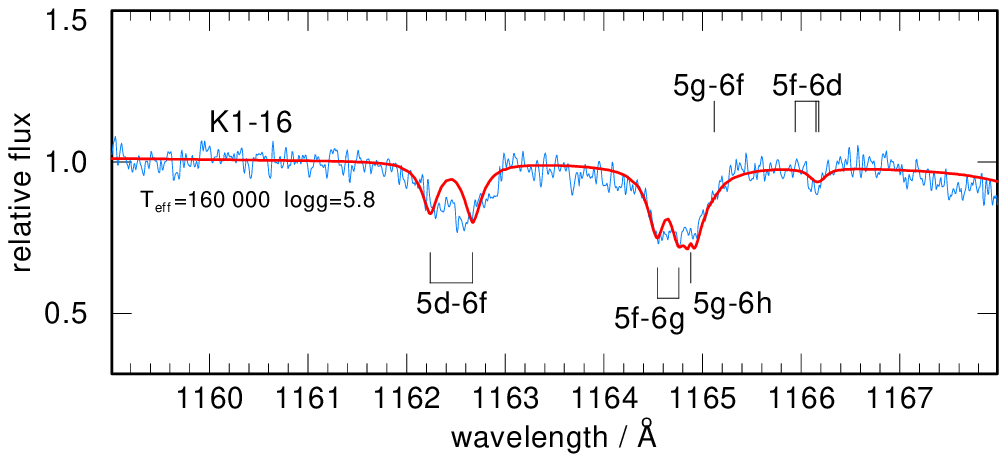}
\caption{\label{fig:ne7} \label{fig:ne8} \emph{Left:} First identification of the
  \ion{Ne}{vii}~973.3~\AA\ line, shown here in the FUSE
  spectrum of the PG1159 star PG1520+525. This strong absorption feature
  is seen in the spectra many hot post-AGB stars, but remained
  unidentified for some years \cite{werner:04}. \emph{Right:}  Discovery of \ion{Ne}{viii} lines in the FUSE
  spectrum of the PG1159-type central star of K\,1-16. This is the first
  detection of \ion{Ne}{viii} in any photospheric spectrum \cite{werner:07b}. Lines from this
  ion are only exhibited by the very hottest post-AGB stars (\Teff$\geq$140\,000~K).}
\end{figure}

\begin{figure}[t]
\includegraphics[width=0.5\textwidth]{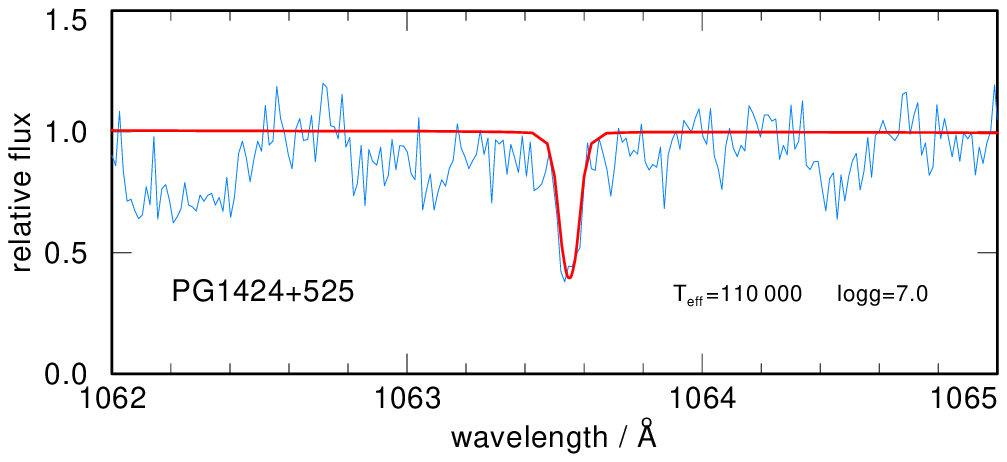}
\includegraphics[width=0.5\textwidth]{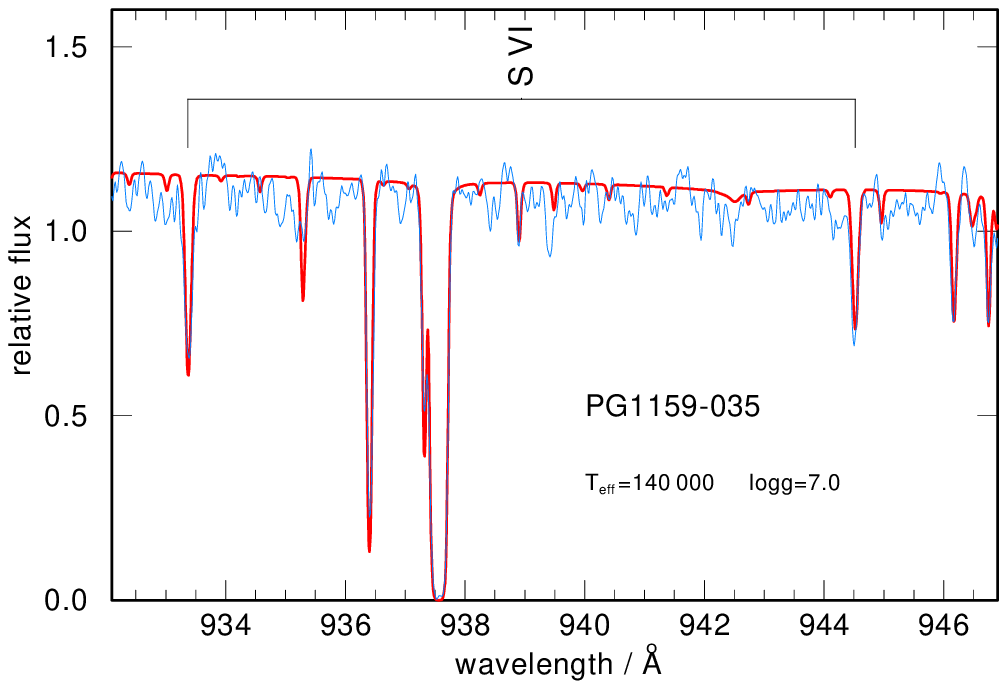}
\caption{\label{fig:ar}\label{fig:s6} \emph{Left:} Discovery of the \ion{Ar}{vii}~1063.55~\AA\ line in the FUSE
  spectrum of the PG1159 star PG1425+535. This is the first
  detection of \ion{Ar}{vii} in the photospheric spectrum  of hot a post-AGB star
  \cite{werner:07a}. \emph{Right:}  FUSE enabled the first abundance determination of sulfur in
  PG1159 stars. Shown here is the \ion{S}{vi} resonance doublet in the
  prototype of the PG1159 spectral class, PG1159$-$035 \cite{jahn:07}.}
\end{figure}

\begin{figure}[htb]
\includegraphics[width=0.5\textwidth]{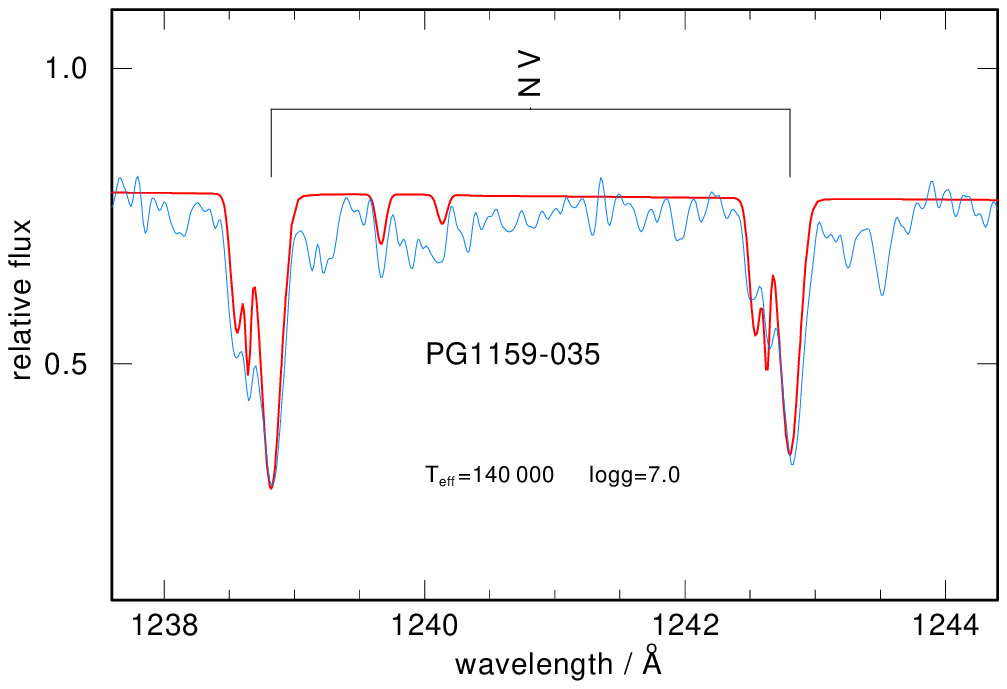}
\includegraphics[width=0.5\textwidth]{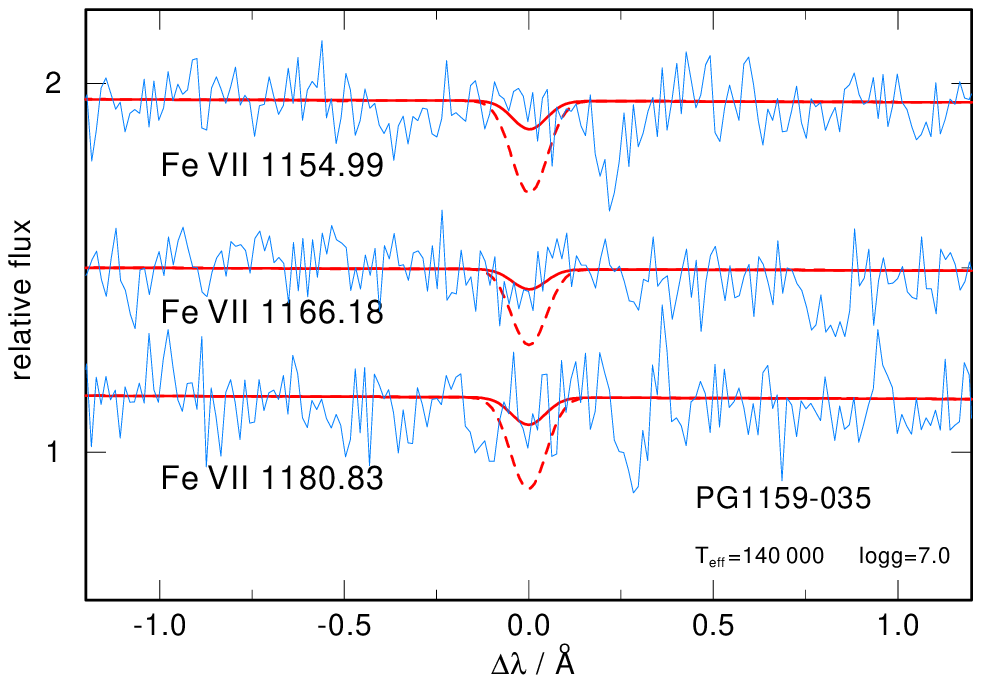}
\caption{\label{fig:n5}\label{fig:fe} \emph{Left:} The high spectral
  resolutionof HST/STIS allows us to distinguish the photospheric \ion{N}{v} resonance
  doublet from the weak blueshifted ISM components. This enabled the first reliable
  N abundance determination in the prototype PG1159-035
  \cite{jahn:07}. \emph{Right:} From the absence of \ion{Fe}{vii} lines in the FUSE
  spectrum a Fe deficiency is concluded. Model spectra were computed with solar and 0.1 solar
  Fe \cite{jahn:07}. }
\end{figure}

A number of chemical elements could be identified
(F, P, S, Ar). In addition, very high ionisation stages of several
elements, which were never seen before in stellar photospheric spectra,
could be identified in the UV spectra for the very first time
(e.g. \ion{Si}{v}, \ion{Si}{vi}, \ion{Ne}{viii}). To
illustrate this, we display in Figs.~\ref{fig:psifn}--\ref{fig:n5}
details of FUSE and HST spectra of PG1159 stars of particularly
interesting wavelength regions, together with synthetic line profile fits.

\noindent
\emph{Hydrogen --} Four PG1159 stars show residual H with an abundance
of 0.17. These objects are the outcome of an AFTP. All other PG1159
stars have H $\lappr$ 0.1 and, hence, should be LTP or VLTP objects.

\noindent
\emph{Helium, carbon, oxygen --} These are the main constituents of
PG1159 atmospheres. A large variety of relative He/C/O abundances is
observed. The approximate abundance ranges are: He=0.30--0.85,
C=0.15--0.60, O=0.02--0.20. The spread of abundances might be explained
by different numbers of thermal pulses during the AGB phase.

\noindent
\emph{Nitrogen --} N is a key element that allows us to decide if the star
is the product of a VLTP or a LTP. Models predict that N is diluted
during an LTP so that in the end N=0.1\%. This low N abundance is
undetectable in the optical and only detectable in extremely good UV
spectra. In contrast, a VLTP produces nitrogen (because of H-ingestion
and burning) to an amount of 1\% to maybe a few percent. N abundances of
the order 1\% are found in some PG1159 stars, while in others it is
definitely much lower.

\noindent
\emph{Neon --} Ne is produced from $^{14}$N that was produced by CNO
burning. In the He-burning region, two $\alpha$-captures transform
$^{14}$N to $^{22}$Ne. Stellar evolution models predict Ne=0.02 in the
intershell. A small spread is expected as a consequence of different
initial stellar masses. Ne=0.02 was found in early optical analyses of a
few stars and, later, in a much larger sample observed with FUSE
\cite{werner:04}.

\noindent
\emph{Fluorine --} F was discovered by
\cite{werner:05} in hot post-AGB stars; in PG1159 stars as well as
H-normal central stars. A strong absorption line
located at 1139.5~\AA\ remained unidentified until we found that it
stems from \ion{F}{vi}. The abundances derived for PG1159 stars show a
large spread, ranging from solar to up to 250 times solar. This was
surprising at the outset because $^{19}$F, the only stable F isotope, is
very fragile and easily destroyed by H and He. A comparison with AGB
star models of \cite{lugaro:04a}, however, shows that such high F
abundances in the intershell can indeed be accumulated by the reaction
$^{14}$N($\alpha$,$\gamma$)$^{18}$F($\beta^+$)$^{18}$O(p,$\alpha$)$^{15}$N($\alpha$,$\gamma$)$^{19}$F,
the amount depending on the stellar mass. We find a good agreement between
observation and theory. Our results also suggest, however, that the F
overabundances found in AGB stars \cite{jorissen:92} can only be
understood if the dredge-up of F in the AGB stars is much more efficient
than hitherto thought.

\noindent
\emph{Silicon --} The Si abundance in evolution models remains almost
unchanged. This is in agreement with the PG1159 stars for which we could
determine the Si abundance.

\noindent
\emph{Phosphorus --} Systematic predictions from evolutionary model
grids are not available; however, the few computed models show P
overabundances in the range 4--25 times solar (Lugaro priv. comm.). This
is at odds with our spectroscopic measurements for two PG1159 stars,
that reveal a solar P abundance.

\noindent
\emph{Sulfur --} Again, model predictions are uncertain at the
moment. Current models show a slight (0.6 solar) underabundance. In
strong contrast, we find a large spread of S abundances in PG1159 stars,
ranging from solar down to 0.01 solar.

\noindent
\emph{Argon --} This element has been identified very recently for the
first time in hot post-AGB stars and white dwarfs \cite{werner:07a}. Among them is one
PG1159 star for which a solar Ar abundance has been determined (Fig.\,\ref{fig:ar}). This is
in agreement with AGB star models which predict that the Ar abundance
remains almost unchanged.

\noindent\emph{Lithium --} Unfortunately, PG1159 stars are too hot to exhibit Li
lines because Li is completely ionised. If Li were detected then it must
have been produced during a VLTP. The discovery of Li in Sakurai's star
is a strong additional hint that it underwent a VLTP and not an LTP.

\noindent
\emph{Iron and Nickel --} \ion{Fe}{vii} lines are expected to be the
strongest iron features in PG1159 stars. They are located in the UV
range. One of the most surprising results is the non-detection of these
lines in three PG1159 stars examined (K1-16, NGC\,7094, PG1159-035; see,
e.g., Fig.\,\ref{fig:fe}). The
derived upper abundance limits (e.g., \cite{werner:03,jahn:07}) indicate that
iron is depleted by about 0.7--2 dex, depending on the particular
object. Iron depletions were also found for the PG1159-[WC] transition
object Abell~78 as well as for several PG1159 progenitors, the [WC]
stars. Such high Fe depletions are not in agreement with current AGB
models. Destruction of $^{56}$Fe by neutron captures is taking place in
the AGB star intershell as a starting point of the s-process; however,
the resulting depletion of Fe in the intershell is predicted to be small
(about 0.2~dex). It could be that additional Fe depletion can occur
during the late thermal pulse. In any case, we would expect a
simultaneous enrichment of nickel, but up to now we were unable to
detect Ni in PG1159 stars at all. While the solar Fe/Ni ratio is about
20, we would expect a ratio close to the s-process quasi steady-state
ratio of about 3. Fittingly, this low ratio has been found in Sakurai's
(cool) LTP object.

\noindent
\emph{Trans-iron elements --} The discovery of s-process elements in
PG1159 stars would be highly desirable. However, this is at present
impossible due to the lack of atomic data. From the ionization
potentials we expect that these elements are highly ionised like iron,
i.e., the dominant ionization stages are \ion{}{vi}~--\ion{}{ix}. To our
best knowledge, there are no laboratory measurements of so highly
ionised s-process elements that would allow us to search for atomic
lines in the observed spectra. Such measurements would be crucial to
continue the element abundance determination beyond the current state.

\section{Conclusions}

It has been realized that PG1159 stars exhibit intershell matter on
their surface, which has probably been laid bare by a late final thermal
pulse. This provides the unique opportunity to study directly the result
of nucleosynthesis and mixing processes in AGB stars. Spectroscopic
abundance determinations of PG1159 photospheres are in agreement with
intershell abundances predicted by AGB star models for many elements
(He, C, N, O, Ne, F, Si, Ar). For other elements, however, disagreement
is found (Fe, P, S) that points at possible weaknesses in the
evolutionary models.

\acknowledgement I thank the conference organizers for travel support and
particularly for
their
kind hospitality and assistance.

\index{paragraph} 

\printindex
\end{document}